\documentclass{elsart}
\usepackage{amsfonts, amsmath}

\newcommand{\eq}[1]{eq.~(\ref{#1})}
\def\c {\mathrm{Const.}}
\def\dist {\operatorname{dist}}

\def\E {{\mathbb E}}
\def\iu {\mathrm i}

\def\liminf {\mathop{\underline{\rm lim}}}

\def\P{\operatorname{Prob}}

\begin{document}

\date{Jan. 16, 2000}
\begin{frontmatter}
\title{Constructive Fractional-Moment Criteria for
Localization in Random Operators } 
\author[Prnctn-Phy,Prnctn-Math]{Michael Aizenman \thanksref{NSF-MA}}
\author[Prnctn-Math]{Jeffrey H. Schenker \thanksref{NSF-JS}}
\author[Zurich]{Roland M. Friedrich}
\author[Caltech]{Dirk Hundertmark\thanksref{DH}}

\address[Prnctn-Phy]{Physics Department,
Princeton University, Jadwin Hall, Princeton, NJ 08544}
\address[Prnctn-Math]{Mathematics Department,
Princeton University, Fine Hall, Princeton, NJ 08544}
\address[Zurich]{Student, Theoretische Physik,
ETH-Z\"urich, CH--8093, Switzerland}
\address[Caltech]{Department of Mathematics 253-37,
Caltech, Pasadena, CA 91125 }
\thanks[NSF-MA]{Supported in part by the NSF Grant PHY-9971149.}
\thanks[NSF-JS]{Supported by an NSF Graduate Research Fellowship.}
\thanks[DH]{Supported by the Deutsche Forschungsgemeinschaft
Grant  Hu 773/1-1.}

\begin{abstract}
We present a  family of finite-volume criteria which cover the
regime of exponential decay for the fractional moments of Green
functions of operators with random potentials.
Such decay is a technically convenient characterization of
localization for it is known to imply spectral localization,
absence of level repulsion, dynamical localization and a
related condition which plays a significant role in the
quantization of the Hall conductance in two-dimensional Fermi gases.
The constructive criteria
also preclude fast power-law decay of the Green
functions at mobility edges.

\end{abstract}


\end{frontmatter}

\section{Introduction}

In the study of Anderson localization the analogy with the 
statistical mechanics of spin systems has often served as 
a  source of  insight  \cite{AALR,MF}.  
In this note we report on some recent results \cite{ASFH}
which were motivated in part by this analogy, and in part
by the desire to develop elementary methods for the study
of different aspects of the localization phenomena.
Our goal is to present only an outline --
the detailed statements and proofs are given in ref.~\cite{ASFH}.

The subject of our discussion are  
random operators acting on the Hilbert space $\ell^2(\Zset^d)$ of
square summable functions defined over the regular $d$-dimensional
lattice $\Zset^d$. A prototypical example is the discrete
Schr\"odinger operator:
\begin{equation}
H_{\omega} \ = \ -\Delta  \ + \ V_{\omega}(x) \; ,
\end{equation}
where $V_{\omega}(x)$ is a random potential, and $\Delta$ is the
nearest neighbor difference operator (``discrete Laplacian''). The
operator may be augmented by the addition of a magnetic field,
and/or random off-diagonal hopping terms.  However, for our
results we assume translation invariance at least in the
stochastic sense and up to gauge transformations (i.e., magnetic
shifts). (The statements may also be adapted to periodic and
quasi-periodic structures.) It is now well understood that such
operators have energy regimes in which the spectrum consists of an
infinite collection of eigenvalues associated to exponentially
localized eigenfunctions. It is also expected, although the theory
is sorely lacking, that in certain situations such operators also
possess energy regions associated to extended states.
 
A very useful tool is provided by
the Green function of $H_\omega$:
\begin{equation}
G_\omega(x,y;E+\iu \eta) \ := \ < x | {1 \over H_\omega - E - \iu
\eta } | y > \; .
\end{equation}
To illustrate the relation with spin systems
one may note the analogy between $G_\omega$ and the spin-spin correlation
functions suggested by the functional integral expression for the former as a
Gaussian integral:
 \begin{equation}
G_\omega(x,y;E+ \iu \eta) \ = \  (-\iu ) {\int [{\mathcal D} \Phi]
    \, \e^{- \iu \left < \Phi, (H_{\omega} - E - \iu
    \eta) \Phi \right >  } \, \Phi(x) \Phi(y)
    \over
    \int [{\mathcal D} \Phi] \,
    \e^{- \iu \left < \Phi, (H_{\omega} - E - \iu
    \eta) \Phi \right >} }\; .
\end{equation}
The relevant rigorous methods familiar from the study of spin
systems do not apply here since the integral involves a complex
action. Nevertheless, exponential bounds for $G_\omega$ in the
limit $\eta \rightarrow 0$ are a ``signature'' of the localized
regime, much as exponential decay of spin-spin correlations
indicates the high-temperature regime.

Since the Green function depends on the disorder, it is
tempting to consider  the
averaged  function $\E \left ( |G_\omega(x,y;E+\iu \eta)|
\right )$, where $\E(\cdot )$ indicates the expectation with
respect to the potential. However, for $E$ in the spectrum of
$H_\omega$, this quantity may diverge as $\eta \rightarrow 0$. As
was realized in ref. \cite{AM}, we can
avoid this problem by considering a ``fractional moment'' of the
Green function: $\E \left ( |G_\omega(x,y;E+\iu \eta)|^s \right )$
for $s < 1$.

Thus, a technically convenient ``signature'' of localization is
the exponential decay of such fractional moments, at some suitable
$s\in (0,1)$,
\begin{equation}
\E(|<x| {1 \over H_{\omega}-E - \iu \eta } | y>|^s) \ \le \ A(s) \
\e^{-\mu(s) |x-y|}  \ , \label{eq:fm}
\end{equation}
where the bound is satisfied uniformly in $\eta \in \Rset$
 for all energies in some range $E\in (a,b)$.

Before we turn to the new results, which offer constructive
criteria for the  validity of the fractional moment condition
(\ref{eq:fm}), let us list several  known implications of this
condition, each of which also presumes some mild regularity
conditions on the distribution of the random potential:

\begin{itemize}
\item[i.]  {\it Spectral localization
(\cite{AM} - using \cite{SiWo}):} The spectrum of $H_{\omega}$
within the interval $(a,b)$ is almost-surely of the pure-point
type, and the corresponding eigenfunctions are exponentially
localized.

\item[ii.] {\it Dynamical localization (\cite{Ai94}):}
Wave packets with energies in the specified range do not spread
(and in particular the {\em SULE} condition of \cite{SULE} is
met):
\begin{equation} \E\left( \sup_{t\in \Rset} |<x| \ \e^{-\iu tH} P_{H\in
(a,b)}\ |y>| \right ) \ \le \widetilde A \e^{-\tilde \mu |x-y|} \;
. \label{eq:dyn}
\end{equation}

\item[iii.] {\it Absence of level repulsion (\cite{Min}).}
Minami  has shown that (\ref{eq:fm}) implies that in the range
$(a,b)$ the energy gaps have Poisson-type statistics.

\item[iv.] {\it Exponential decay
of the projection kernel (\cite{AG})}:
\begin{equation}
\E(\, |<x|\ P_{H\le E}\ |y>|\, )  \ \le \widehat A \e^{-\hat \mu
|x-y|} \; . \label{eq:projection}
\end{equation}
This condition plays an important role in the quantization of Hall
conductance in the ground state of the two--dimensional electron
gas with Fermi level $E_{F}\in (a,b)$ \cite{AS2,BES,AG}.

\end{itemize}

The fractional moment condition (\ref{eq:fm}) has already been
established for certain regimes: large disorder and extreme
energies \cite{AM}, and also at weak disorder but for energies far
from the spectrum of $-\Delta$ \cite{Ai94}. Our
new results extend the reach of the fractional
moment method by showing that the entire region in which
(\ref{eq:fm}) holds can, in principle, be determined by a sequence
of  finite calculations.

In their general appearance, these results may remind one of some
of the constructive criteria for the high temperature phases in
certain models of statistical mechanics, which are mentioned
below. As in that case, the results also yield  some conclusions
about the critical behavior, which in the present context refers
to the behaviour in the vicinity of the mobility edge -- wherever
such an edge occurs.

Before the introduction of the fractional-moment method,
localization regimes have been established using the
multiscale analysis of Fr\"ohlich and Spencer \cite{FS} which
yields exponential bounds of the form:
\begin{equation}
|G_\omega(x,y;E) | \ \le \ A(\omega, x) \e^{-\mu(\omega) |x-y|} \; ,
\end{equation}
with $\mu(\omega) > 0 $ and $A(\omega, x) < \infty$ for almost
every $\omega$ and all $x\in \Zset^d$.  The bounds which the
multiscale analysis provides for the  probability of the
exceptional cases decay  faster than any power of $|x-y|$ but not
exponentially fast. It is difficult to use such results to
demonstrate exponential bounds on {\em expectation values} such as
those seen in equations  (\ref{eq:projection}) and (\ref{eq:dyn}).
Nevertheless we note that dynamical localization was recently
established also by arguments starting from the bounds provided by
the multiscale analysis \cite{DS99},  using methods not related to
this work. We shall return to the relation between the
fractional-moment method and the multiscale analysis in the final
section of this note.

\section{The finite-volume criteria}

The results presented herein describe certain conditions which
when satisfied by the operator $H_{\Lambda;\omega}$ obtained by
restricting $H_\omega$ to some finite volume $\Lambda$ are sufficient
to deduce the fractional moment condition (\ref{eq:fm}) for the
full operator $H_\omega$. For simplicity we state these results
only in the case of random Schr\"odinger operators.  The reader is
directed to \cite{ASFH} for versions which apply to more general
operators.

To guarantee that the fractional moments are finite, we require
certain regularity of the joint probability distribution of the
site potentials $V(x)$.  An additional technical assumption
related to the ``decoupling lemmas'' used in \cite{AM,Ai94,AG} is
also required.  In their mildest form the conditions required
are somewhat technical to
state, so for the present note  we shall call  a probabilty
distribution  of the potential {\em regular} if the  site values
$V(x)$ are independent
identically distributed random variables whose distribution has a
bounded density with compact support.
The interested reader may find the more general assumptions
in \cite{ASFH}.

In order to state our results, we must introduce some notation.
Given a finite region $\Lambda \subset \Zset^d$, we denote by
$\Gamma(\Lambda)$  the set of lattice bonds (nearest neighbor pairs)
connecting sites in
$\Lambda$ with sites in $\Zset^d \setminus \Lambda$,  and
 by $\Lambda^+$ the region obtained
from $\Lambda$ by adding to it all of its nearest neighbors. The
number of elements of a set $W$ is denoted $|W|$.

As mentioned above,  $\E(\cdot)$ indicates the
expectation with respect to the random potential.
We also let
\begin{equation}
\E_{+ \iu 0 \, (- \iu 0)} (\, G(E) \, ) \ := \
\lim_{\stackrel{ \eta \searrow 0}
{ (\eta \nearrow 0)}}
\E(\, G(E + \iu \eta) \, ) \; .
\end{equation}

Following is the first of our results.
\begin{thm}
\label{thm:1} Let $H_{\omega}$ be a random Schr\"odinger operator
with a regular distribution of the random potential. Then for each
$s < 1$  there exists $C_s < \infty $ such that if for some $E \in
\Rset$ (in fact also $E \in \Cset$) and some finite region
$\Lambda\subset \Zset^d $ which contains the origin $O$:
\begin{equation}
 \left ( 1 + { C_s \over \lambda^s} \
|\Gamma(\Lambda)| \right )^2
   \sum_{<u,u'> \in \Gamma(\Lambda)}
\E \left ( |<O| {1 \over H_{\Lambda;\omega} - E} |u>|^s
            \right ) \ < \ 1 \; ,
\label{eq:cond1}
\end{equation}
then $H_{\omega}$ satisfies the fractional-moment condition
(\ref{eq:fm}), and there exist $\mu(s) > 0, A(s) < \infty $,
which depend on $E$ only  through the value of the LHS
in \eq{eq:cond1},
so that for any region $\Omega \subset \Zset^d$,
\begin{equation}
\E_{\pm \iu 0}\left( |<x| {1 \over H_{\Omega;\omega}-E} |y> |^s
\right)\ \le \ A(s)\  \e^{-\mu(s)\, \dist_{\Omega}(x,y)} \; ,
\label{eq:thm1}
\end{equation}
with
\begin{equation}
\dist_{\Omega}(x,y)= \min\{|x-y|, [\dist(x,\partial \Omega) +
\dist(y,\partial \Omega)] \} \; . \label{eq:dist}
\end{equation}
\end{thm}

The modified metric, $\dist_{\Omega}(x,y)$,  is a distance
function relative to which the entire boundary of $\Omega$ is
regarded as one point. It permits us to state that there is
exponential decay in the ``bulk'' without ruling out the possible
existence of extended {\em boundary states} in some geometry.

One may also formulate finite-volume criteria which rule out
extended boundary states, that is which permit us to conclude
exponential decay of the fractional moments in {\em any} region
$\Omega$.  The trade-off is that the finite volume test, presented
in the next result,  is a bit more involved.

\begin{thm}
\label{thm:2} Let $H_{\omega}$ be a random Schr\"odinger operator
with a regular distribution of the random potential. Then for each
$s < 1$ there exists $\widetilde C_s > 0$ such that if for some
$E\in \Rset$ (alternatively, complex $E$) and some finite region
$O \in \Lambda \subset \Zset^d$:
\begin{equation}
 \max_{W \subset \Lambda} \ \left\{ |\Gamma(\Lambda^+)| \
{\widetilde C_s \over \lambda^s} \ \sum_{<u,u'>
\in
\Gamma(\Lambda) } \E\left(|<O| {1 \over H_{W ; \omega} - E}
|u>|^s
\right) \right\}  \  < \ 1 \; , \label{eq:cond2}
\end{equation}
then there are $\mu(s) > 0$ and $A(s) < \infty $ --- which depend
on the energy $E$ only through the value of the LHS of
\eq{eq:cond2}
--- such that for any region $\Omega \subset \Zset^d$
\begin{equation}
\label{eq:thm2}
 \E_{\pm \iu
0}\left( |<x| {1 \over  H_{\Omega;\omega} - z } |y> |^s \right)\
\le \ A(s) \  \e^{-\mu(s) \, |x-y| } \; .
\end{equation}
\end{thm}

It is rather obvious that the collection of finite-volume criteria
provided in  Theorem~\ref{thm:2} covers the entire regime in which
the conclusion, \eq{eq:thm2}, holds.  The corresponding statement
for Theorem~\ref{thm:1} is a bit less immediate, but it is also
true:

\begin{thm}
Let $H_{\omega}$ be a random Schr\"odinger operator with a regular
distribution of the random potential, and fix $s < 1$.
If at some energy $E$ (or $E\in \Cset$)
the localization condition (\ref{eq:fm}) is satisfied, with some
$A< \infty$ and $\mu> 0$, then for all  large enough (but finite)
$L $ the condition (\ref{eq:cond1}) is met for $\Lambda =
[-L,L]^d$. \label{thm:3}
\end{thm}

\section{Some Implications}

We shall now mention a number of implications of the finite-volume
criteria for fractional moment localization.

First, of course, are explicit bounds, and we already obtain such
bounds with a single site estimate corresponding to $ \Lambda =\{
0 \} $.  The test provided by Theorem~\ref{thm:2} is met for all
$\lambda$ and $E$ such that:
\begin{equation}
{2 d^2 (2d+1) \, C_s  \over \lambda^s } \
\E(\, {1 \over |\lambda V - E|^s} \, ) \  \ < \ \ 1.
\end{equation}
This implies localization for strong disorder, and at extremal
energies, in the manner of ref.~\cite{AM}.

The above explicit criterion  may now be systematically improved.
However, since the calculations quickly become quite laborious,
perhaps the main benefit are certain qualitative statements. Those
bear some resemblance to results derived using the multiscale
approach; however the conclusions drawn here go beyond the latter
by yielding results on the exponential decay of the {\em mean
values}.

\subsection{Fast power decay $\Rightarrow$ exponential decay}

An interesting and useful implication (as seen below) is that fast
enough power law implies exponential decay. In this sense, random
Schr\"odinger operators join other statistical mechanical models
in which such principles have been previously recognized. The list
includes the general Dobrushin-Shlosman results~\cite{DoSh}  and
the more specific two-point function bounds for: percolation
\cite{Ham,AiNew}, Ising ferromagnets \cite{Simon,Lieb}, certain
$O(N)$ models \cite{AiSi}, and time-evolution models
\cite{AiHo,MaSh}.

\begin{thm}
Let $H_{\omega}$ be a random Schr\"odinger operator on
$\ell^2(\Zset^d)$ with a regular potential. Then there are $L_o,
B_1, B_2 <\infty$ such that if  for some $E\in \Rset$ and some
finite $L \ge L_o $,  either
\begin{equation}
         \sup_{\ L /
2 \le \|y\| \le L}
            \E \left(     |<O| {1 \over
H_{\Lambda_L, \omega} - E} |y>|^s
            \right ) \ \le \ B_1 /   L^{3(d-1)} \; ,
\label{eq:powerlaw1}
\end{equation}
or
\begin{equation}
  \sup_{\ L / 2 \le \|y\| \le L}
      \E \left( |<O| {1 \over
                    H_{\omega} - E } |y>|^s
\right ) \ \le \ B_2 / L^{4(d-1)}  \; , \label{eq:powerlaw2}
\end{equation}
where $\Lambda_L=[-L,L]^d$ and $\|y\| \equiv \sum_{j} |y_j|$, then
the {\em exponential} localization (\ref{eq:fm}) holds for all
energies in some open interval $(a,b)$ containing $E$.
\label{thm:power=>exp}
\end{thm}

\subsection{Lower bounds for $G_{\omega}(x,y; E_{{\rm edge} }+i0)$
at mobility edges}

Boundary points of the continuous spectrum  are  referred to as
{\em mobility edges}.  The random Schr\"odinger operators
considered here are ergodic, hence the location of such points
does not depend on the realization \cite{KuSu}. However, except
for the Bethe lattice~\cite{K1}, the proof of the occurrence of
continuous spectrum is still an open problem. Nevertheless, it is
interesting to note that Theorem~\ref{thm:power=>exp} yields the
following pair of lower bounds on the decay rate of the Green
function at mobility edges, $E_{{\rm edge} }$, for a random
Schr\"odinger operator with regular potential:
\begin{equation}
    \ \sup_{\
L / 2 \le \|y\| \le L}
      \E \left(
                |<O| {1 \over
H_{[-L,L]^d; \omega} - E_{{\rm edge} } } |y>|^s
            \right ) \ \ge
\ B_1 \  L^{-3(d-1)} \; ,
\label{eq:mobilityedge1}
\end{equation}
and
\begin{equation}
    \ \sup_{\ L / 2 \le \|y\| \le L}
      \E \left(
|<O| {1 \over H_{\omega} - E_{{\rm edge} } } |y>|^s
            \right ) \ \ge \ B_2 \  L^{-4(d-1)} \; .
\label{eq:mobilityedge2}
\end{equation}
We do not expect these bounds to be optimal.

Vaguely similar bounds are known for the critical two-point
functions in the statistical mechanical models mentioned
above.

\subsection{Extending off the real axis}

The following statement is of somewhat technical interest, but it
has interesting implications, such as the decay of the projection
kernel, for which it is useful to have bounds on the resolvent at
$E+i\eta$ which are uniform in $\eta$. Such bounds permit
integrating the resolvent estimates along contours which cut the
real axis, as in the derivation of (\eq{eq:projection}) in
ref.~\cite{AG}.

\begin{thm}  Let $H_{\omega}$ be a random Schr\"odinger operator with
a regular potential. Suppose that for some $E\in \Rset$, and
$\Delta E > 0$, the following bound holds uniformly for $\xi \in
[E-\Delta E, E+\Delta E ]$:
\begin{equation}
    \E \left(  |<x| {1 \over H_{\omega} - \xi - i 0}
|y>|^s
            \right ) \ \le \ A\,  e^{-\mu |x-y|}  \; .
\label{eq:real}
\end{equation}
Then for all $\eta \in \Rset$:
\begin{equation}
    \E \left(  |<x| {1 \over H_{\omega} - E - i \eta}
|y>|^s
            \right ) \
 \le \ \widetilde A\ \ e^{-\tilde \mu  |x-y|}
\; ,
\label{eq:eq:strip}
\end{equation}
with some $\widetilde A  < \infty$ and $\tilde \mu > 0$ -- which
depend on  $\Delta E$ and  the bound (\ref{eq:real}).
\label{thm:strip}
\end{thm}

\subsection{Localization in spectral tails.}

The finite volume criteria presented above allow us to conclude
exponential localization from suitable bounds on the density of
states of the operators in regions $\Lambda_{L} = [-L,L]^d$.  The
following statement will be useful for such a purpose.

\begin{thm} Let $H_{\omega}$ be a random Schr\"odinger operator on
$\ell^2(\Zset^d)$ with a regular potential.  For each $L>0$ there
exist $\delta_L > 0$ and $P_L > 0$ such that if
\begin{equation}
\P \left [ \dist \left ( \sigma(H_{\Lambda_{L}; \omega}), E
\right) \le \delta_L \right ] \ < \ P_L \; ,
\label{eq:finitecondition}
\end{equation}
then the exponential localization condition (\ref{eq:fm}) holds in
some open interval containing $E$.  Furthermore, given $\beta \in
(0,1)$ and $\xi > 3 (d-1)$, it is possible to choose $\delta_L$
and $p_L$ such that $  \liminf L^\beta \delta_L > 0$ and $ \liminf
L^\xi P_L > 0$.
\label{thm:density}
\end{thm} 

\noindent{\bf Remarks:} 1. It is of interest to combine the
criterion presented above with Lifschitz tail estimates on the
density of states at the bottom of the spectrum and at band edges.
As an example, consider the bottom of the spectrum of $H_\omega$:
$E_0 = - \lambda V_0$ where $V_0$ is the minimum value in the
support of $V$.  Using Lifschitz tail estimates, it is possible to
show that \cite{Si85}:
\begin{equation}
\P \left [ \inf \sigma(H_{\Lambda_{L}; \omega}) \le E_0 + \Delta E
\right ] \ \le \ \c \ L^d e^{-\Delta E^{-d/2}} \; .
\end{equation}
By choosing $\Delta E \propto L^{-\beta}$ with $\beta \in (0,1)$
for large enough $L$, we infer fractional moment localization from
this bound via Theorem~\ref{thm:density}. Previous results in this
vein may be found in \cite{BCH,FiKl,KSS,Klopp99,Stollmann}.

2. The input conditions (\ref{eq:finitecondition})
are similar to the input used in the
multiscale analysis.  In fact, there it is not important that $\xi
> 3(d-1)$, and it suffices to assume the condition is met
for some $\xi >0$. However, one may note that wherever the multiscale
analysis applies, its conclusion allows to deduce the
condition as stated here. Thus, the exponential localization in
the stronger sense discussed in our work may be concluded
also for the regime for which localization may be established
through the multiscale analysis.


\begin{thebibliography}{10}

\bibitem{AALR}
E.~Abraham, P.~W. Anderson, D.~C. Licciardello, and T.~V. Ramakrishnan,
  ``Scaling theory of localization: {A}bsence of quantum diffusion in two
  dimensions,'' {\em Phys. Rev. Lett.},  {\bf 42},  673, (1979).

\bibitem{MF}
A.~D. Mirlin and Y.~V. Fyodorov, ``Localization transition in the {A}nderson
  model on the {B}ethe lattice: spontaneous symmetry breaking and correlation
  function,'' {\em Nucl. Phys. B},  {\bf 366},  507, (1991).

\bibitem{ASFH}
M.~Aizenman, J.~Schenker, R.~Friedrich, and D.~Hundertmark, ``Finite-volume
  criteria for {A}nderson localization,'' (1999 preprint). \\ 
\newblock http://xxx.lanl.gov/abs/math-ph/9910022 

\bibitem{AM}
M.~Aizenman and S.~Molchanov, ``Localization at large disorder and at extreme
  energies: an elementary derivation,'' {\em Comm. Math. Phys.},  {\bf 157},
  245, (1993).

\bibitem{SiWo}
B.~Simon and T.~Wolff, ``Singular continuous perturbation under rank one
  perturbation and localization for random {H}amiltonians,'' {\em Commun. Pure
  Appl. Math.},  {\bf 39},  75, (1986).

\bibitem{Ai94}
M.~Aizenman, ``Localization at weak disorder: some elementary bounds,'' {\em
  Rev. Math. Phys.},  {\bf 6},  1163, (1994).

\bibitem{SULE}
R.~del Rio, S.~Jitomirskaya, Y.~Last, and B.~Simon, ``Operators with singular
  continuous spectrum. {I}{V}. {H}ausdorff dimensions, rank one perturbations,
  and localization,'' {\em J. Anal. Math.},  {\bf 69},  153, (1996).

\bibitem{Min}
N.~Minami, ``Local fluctuation of the spectrum of a multidimensional {A}nderson
  tight binding model,'' {\em Comm. Math. Phys.},  {\bf 177},  709, (1996).

\bibitem{AG}
M.~Aizenman and G.~M. Graf, ``Localization bounds for an electron gas,'' {\em
  J. Phys. A: Math. Gen.},  {\bf 31},  6783, (1998).

\bibitem{AS2}
J.~E. Avron, R.~Seiler, and B.~Simon, ``Charge deficiency, charge transport and
  comparison of dimensions,'' {\em Comm. Math. Phys.},  {\bf 159},  399,
  (1994).

\bibitem{BES}
J.~Bellissard, A.~van Elst, and H.~{Schulz-Baldes}, ``The noncommutative
  geometry of the quantum {H}all effect,'' {\em J. Math. Phys.},  {\bf 35},
  5373, (1994).

\bibitem{FS}
J.~Fr{\"o}hlich and T.~Spencer, ``Absence of diffusion in the {A}nderson tight
  binding model for large disorder or low energy,'' {\em Comm. Math. Phys.},
  {\bf 88},  151, (1983).

\bibitem{DS99}
D.~Damanik and P.~Stollmann, ``Multi-scale analysis implies strong dynamical
  localization,'' (1999 preprint).
\newblock http://xxx.lanl.gov/abs/math-ph/9912002 

\bibitem{DoSh}
R.~L. Dobrushin and S.~B. Shlosman, ``Completely analytical interactions:
  constructive description,'' {\em J. Statist. Phys.},  {\bf 46}, no.~5-6,
  983--1014, (1987).

\bibitem{Ham}
J.~M. Hammersley, ``Percolation processes {II}. the connective constant,'' {\em
  Proc. Camb. Phil. Soc.},  {\bf 53},  642, (1957).

\bibitem{AiNew}
M.~Aizenman and C.~M. Newman, ``Tree graph inequalities and critical behavior
  in percolation models,'' {\em J. Stat. Phys.},  {\bf 36},  107, (1984).

\bibitem{Simon}
B.~Simon, ``Correlation inequalities and the decay of correlations in
  ferromagnets,'' {\em Comm. Math. Phys.},  {\bf 77}, no.~2,  111, (1980).

\bibitem{Lieb}
E.~H. Lieb, ``A refinement of {S}imon's correlation inequality,'' {\em Comm.
  Math. Phys.},  {\bf 77}, no.~2,  127, (1980).

\bibitem{AiSi}
M.~Aizenman and B.~Simon, ``Local {W}ard identities and the decay of
  correlations in ferromagnets,'' {\em Comm. Math. Phys.},  {\bf 77}, no.~2,
  137, (1980).

\bibitem{AiHo}
M.~Aizenman and R.~Holley, ``Rapid convergence to equilibrium of stochastic
  {I}sing models in the {D}obrushin {S}hlosman regime,'' in {\em Percolation
  theory and ergodic theory of infinite particle systems (Minneapolis, Minn.,
  1984--1985)},  1, New York: Springer, 1987.

\bibitem{MaSh}
C.~Maes and S.~B. Shlosman, ``Ergodicity of probabilistic cellular automata: a
  constructive criterion,'' {\em Comm. Math. Phys.},  {\bf 135}, no.~2,  233,
  (1991).

\bibitem{KuSu}
H.~Kunz and B.~Souillard, ``Sur le spectre des op{\'e}rateurs aux
  diff{\'e}rences finies al{\'e}atoires,'' {\em Comm. Math. Phys.},  {\bf 78},
  no.~201, (1980).

\bibitem{K1}
A.~Klein, ``Absolutely continuous spectrum in the {A}nderson model on the
  {B}ethe lattice,'' {\em Mathematical Research Letters},  {\bf 1},  399,
  (1993).

\bibitem{Si85}
B.~Simon, ``{L}ifschitz tails for the {A}nderson model,'' {\em J. Stat. Phys.},
   {\bf 38},  65, (1985).

\bibitem{BCH}
J.~M. Barbaroux, J.-M. Combes, and P.~D. Hislop, ``Localization near band edges
  for random {S}chr\"odinger operators,'' {\em Helv. Phys. Acta},  {\bf 70},
  16, (1997).
\newblock Papers honouring the 60th birthday of Klaus Hepp and of Walter
  Hunziker, Part II (Z\"urich, 1995).

\bibitem{FiKl}
A.~Figotin and A.~Klein, ``Localization of electromagnetic and acoustic waves
  in random media. lattice models,'' {\em J. Stat. Phys.},  {\bf 76},  985,
  (1994).

\bibitem{KSS}
W.~Kirsch, P.~Stollmann, and G.~Stolz, ``Localization for random perturbations
  of periodic {S}chr\"odinger operators.,'' {\em Rand. Oper. Stoch. Eq.},  {\bf
  6},  241, (1998).

\bibitem{Klopp99}
F.~Klopp, ``Internal {L}ifshits tails for random perturbations of periodic
  {S}chr\"odinger operators.,'' {\em Duke Math. J.},  {\bf 98},  335, (1999).

\bibitem{Stollmann}
P.~Stollmann, ``{L}ifshitz asymptotics via liner coupling of disorder,'' (1999
  preprint).

\end{thebibliography}

\end{document}